\documentclass[12pt]{article}

\usepackage{graphics}
\usepackage{epsfig}
\usepackage{mathrsfs}
\usepackage{multirow}

\newcommand{\nn}{\nonumber}

\newcommand{\bq}{\begin{eqnarray} }
\newcommand{\eq}{\end{eqnarray} }

\def\ga{\mathrel{\raise.3ex\hbox{$>$\kern-.75em\lower1ex\hbox{$\sim$}}}}
\def\la{\mathrel{\raise.3ex\hbox{$<$\kern-.75em\lower1ex\hbox{$\sim$}}}}
\begin{document}\begin{titlepage}

\begin{flushright}
\end{flushright}

\vspace*{1.5cm}
\begin{center}
{\Large {\bf $SU(5)$ unification for Yukawas through SUSY threshold effects\\[-0.05in]
} }

\vspace*{1.5cm}
 {\large {\bf Ts. Enkhbat\footnote{E-mail address:
 enkhbat@ictp.it}}}\\
{\it The Abdus Salam ICTP, \\
Strada Costiera 11, 34014 Trieste, Italy }

\end{center}

 \vspace*{1.5cm}

\begin{abstract}

It is well known that the supersymmetric finite threshold effects
can induce substantial corrections to the Standard Model fermion masses. This opens an alternative
possibility to correct the problematic mass ratios of the lighter
generations within the minimal $SU(5)$ GUT. We
show that with large soft $A$--terms, one can achieve simple
unification for lighter generations without additional Higgs
multiplet, while having sfermions lighter than $1$ TeV. The presence
of such large $A$--terms will distort the sfermion mass spectrum upon
running from GUT scale down to the electroweak scale making it distinct
from the universal SUSY breaking sector, especially in the first two generations. The
implications of these splittings are studied in $K$ and $D$ meson oscillations and
in rare processes $D^+\rightarrow \pi^+\nu\bar{\nu}$ and
$K^+\rightarrow \pi^+\nu\bar{\nu}$, and in the latter case the effect
is found to be important.

\end{abstract}

\end{titlepage}
\newpage
\section{Introduction}

The supersymmetric (SUSY) version of the Standard Model (SM), while having a stabilized
Higgs mass, displays a much better unification of gauge couplings than the non SUSY version. The Yukawa couplings of $b$--quark and
$\tau$--lepton unify at a reasonably good level as well: a slight
discrepancy of $\sim$~20$\%$ can be remedied by various corrections. These successes
fail to extend to the lighter two generations. In
particular, the experimentally determined mass ratio between the  down
and strange quarks is an order of magnitude larger than the electron to muon ratio, which are
predicted to be equal to each other if the minimal unification is
assumed at the scale of the grand unification theory (GUT) for all generations.

In most GUT models this shortcoming is accounted by adding either Yukawa
interactions of a new Higgs multiplet or multiplets such as {\bf 45} in $SU(5)$
\cite{Georgi:1979df} or higher dimensional operators \cite{EmmanuelCosta:2003pu}.
In SUSY theories another possibility of correcting these wrong relations opens up due
to the threshold effects from SUSY breaking which was reported first
time in Ref.~\cite{Buchmuller:1982ye}, and applied for GUT in Ref.~\cite{Hall:1985dx}. Since
then, there have been many studies on these effects of SUSY breaking on
the fermion masses. An incomplete list is given in
Refs.~\cite{Banks:1987iu}--\cite{Antusch:2008tf}. In
Refs.~\cite{Hall:1993gn, Carena:1994bv} the importance of these corrections for b--quark in the large
$\tan\beta$ limit has been pointed out for unified models based on $SO(10)$. In particular, Hall et~al. in
Ref.~\cite{Hall:1993gn} showed that the loop induced $Qd^c
H_u^*$ interaction contributes a large effect to the down--type quark masses due
to $\tan\beta$ enhancement. By now, these SUSY threshold corrections are
integral part of phenomenological studies \cite{Pierce:1996zz} and
every popular code for SUSY spectrum includes them at least for the third generation.

In parallel to these studies there have been many theoretical models
which took the advantage of these corrections to explain the observed fermion mass
hierarchies. In Ref.~\cite{Hempfling:1993kv} a model with 4th family and horizontal gauge
symmetry has been considered where it was shown that one can achieve
unification for all Yukawas without additional Higgs representation
or higher dimensional operators. Along the similar line, in Ref.~\cite{ArkaniHamed:1996zw}
models were proposed where the masses of the
lightest charged fermions are induced purely by the SUSY threshold
corrections through flavor violating soft masses. In Ref.~\cite{Babu:1998tm}
Babu et~al. showed that the CKM mixing can be induced in a similar
manner in left--right models.

All these studies could be classified in the following two categories:
models which (i) explain the fermion mass pattern by soft parameters
or instead (ii) use them to achieve Yukawa unifications for certain GUT models.
In both cases one common trend that has been observed in a number of studies was to
move away from a minimal choice for the SUSY breaking parameters for
achieving acceptable fermion masses and mixings \cite{DiazCruz:2000mn}--\cite{Antusch:2008tf}.
Diaz~Cruz et~al. \cite{DiazCruz:2000mn} has
considered the effect of large flavor violating A--terms for down type
quarks in the first two family. They start with the minimally unified Yukawa
couplings then correct the wrong GUT relation via the large A--term
effect for the down--type quarks. With the form they have chosen for the
A--terms the correct effective Yukawa couplings and the Cabibbo mixing
were obtained. To do so, on the other hand, they have concluded that the GUT relation can be corrected
with sfermions heavier than $4.4$ TeV and $\tan\beta\sim 2$ while gluino is lighter than TeV
to be consistent with the FCNC constraints mostly from $\mu\rightarrow
e\gamma$.

Soon the LHC will start and probe a new energy frontier around TeV. These
corrections become much more interesting if the sfermions are
lighter and hopefully reachable at the LHC. Then, presence of large
A--terms could be related to the observed sfermion masses. This will be complimentary
to any new FCNC signals associated with such corrections. If
breakthroughs happen experimentally on both sides, eventually these
corrections can be tested or ruled out by the experiments. Also very low $\tan\beta\la 3$ seems to be excluded by LEP~II
Higgs search analysis \cite{LEPHiggs} except for very small window for $\tan\beta\la 1$ in the case of no
left--right scalar mixings which gives no radiatively induced
correction to the fermion masses. Thus it is desirable to
study the issue of the minimal unification in the range of moderate
and large $\tan\beta$ range.

In this paper we study large A--terms for the down--type quarks, which are not proportional to
the corresponding Yukawa couplings, nevertheless have flavor diagonal form, that corrects
the wrong GUT ratios. This choice, as we demonstrate, will escape the
FCNC constraints even for sub TeV sfermion masses. Such large A--terms split the
masses of down type sfermions in the first two generations, making the
spectrum distinct from usual universal SUSY parameters which could
be probed at the LHC and/or ILC. We study the $D$ and $K$ meson
oscillations, rare processes $D\rightarrow\pi\nu\bar{\nu}$ and
$K\rightarrow\pi\nu\bar{\nu}$. Although the rate for the $D$--meson is
found to be four order of magnitude larger than its SM prediction, is
still far from the reach of ongoing BESIII experiment. On the other
hand we have found a large effect in the Kaon case in some of our
solutions of the SUSY threshold induced unification. When the
precision of the branching ratio improves \cite{Anelli:2005ju}, it
could be used for determining whether there is a trace of such new physics
through the SM global
fit.

The structure of the paper is as follows. In section 2, I give a brief
review of the finite corrections to the fermion masses, and explain
based on qualitative arguments why we need non minimal soft parameter
for the minimal unification. The section 3 is devoted for the details of our
numerical calculations and main results. The conclusions is given in
section 4. The relevant formulae for the SUSY threshold corrections
to the Yukawa couplings are given in the Appendix.

\section{Radiative Corrections to 
Fermion Masses in the MSSM}

In this section, we review the SUSY threshold corrections to the
fermion masses and highlight the qualitative features of such
corrections. This will enable us to see what choices of soft
parameters would induce the correct level of threshold effects that are needed for changing
the wrong GUT ratio.

\subsection{The SUSY threshold corrections}

\begin{figure}
   \centering
\resizebox{14cm}{!}{\includegraphics{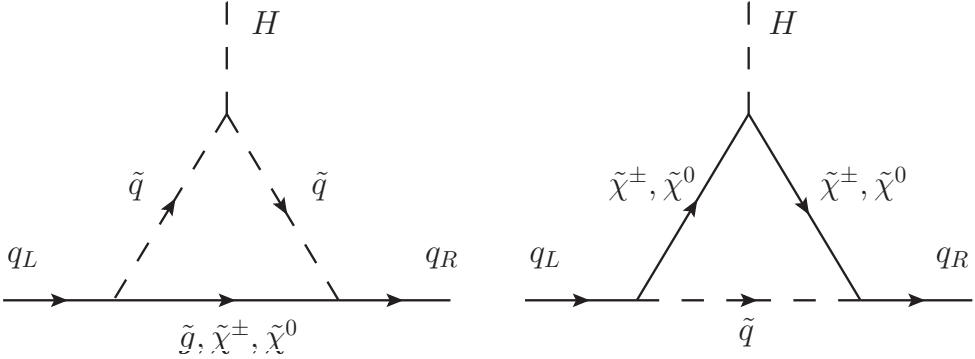}}
\caption{\it The diagram for the finite corrections to the quark
  Yukawa couplings.}
\label{rfm1}
\end{figure}

Whenever a heavy field or fields decouple from a theory they induce
finite shifts in the parameters of the theory. In the SUSY
extension of the SM, these threshold corrections are induced as the SUSY
partners of the SM fields decouple \cite{Buchmuller:1982ye}. The decoupling scale is believed
to be around or not much higher than the electroweak
symmetry breaking scale if SUSY is to stabilize the gauge
hierarchy. In particular, the Yukawa couplings receive finite
threshold corrections from gaugino--sfermion, Higgsino--sfermion loops The corresponding diagrams are depicted in Figure \ref{rfm1} in
the case of quarks. These corrections are especially important due to the fact that the down--type quarks and
charged leptons
obtain a new Yukawa interaction to the up--type Higgs doublet, which, upon
electroweak symmetry breaking, results in $\tan\beta$
enhanced corrections \cite{Hall:1993gn}.

Now we elaborate on the details of these corrections in the case of
quarks.  The full expressions for these corrections are given
in the Appendix, which we have used in our numerical analysis.
The total correction for the quarks are given as follows:
\begin{eqnarray}\label{eq1}
\left(\delta m_q\right)=\left(\delta m^{G}_q\right)+
\left(\delta m^{N}_f\right)+
\left(\delta m^{C}_d\right),
\end{eqnarray}
where the gluino--squark loop induced correction is given by
\begin{eqnarray}\label{eq2}
\left(\delta m^{G}_q\right)_{ij}&\simeq&-\frac{2\alpha_s}{3\pi}(m^q_{LR})m_{\tilde{g}}
I\left( m^2_{\tilde{Q}},m^2_{\tilde{q}^c},m_{\tilde{g}}^2\right),
\end{eqnarray}
The definitions of the mass parameters in the above formula are
given in the Appendix along with the details of the remaining two corrections.
The loop function behaves approximately as $I(m^2,m^2,m^2)\simeq1/(2m^2)$.
To see the qualitative features, here we concentrate  on the
finite correction from  the gluino--squark
loop which is usually the dominant one. For instance, the Bino--squark
induced correction has $8\alpha_s/(3\alpha^\prime)\simeq 0.03$ factor
compared to the above correction.
Here we approximate the mass eigenvalues of the squaks by their soft--mass
parameters as $m^2_{\tilde{q}}\simeq m^2_{\tilde{Q}} \simeq m^2_{\tilde{q}^c}$, ignoring the
mixings as well.
If we assume the A--terms proportional to the corresponding Yukawa couplings the
correction takes the following form
\bq\label{aprTS1}
\frac{\delta
  m_{d_i}}{m_{d_i}^0}\simeq-\frac{\alpha_s}{3\pi}\frac{ m_{\tilde{g}}(a_0-\mu\tan\beta)}{m_{\tilde{d}_i}^2}.
\eq
From this formula we observe that the effect does not decouple in the limit of large SUSY breaking parameters. Also, concentrating on the term proportional to
$\tan\beta$, it is easy to realize that the induced effect could be as large as the
tree level term in the large $\tan\beta$ limit. For example, for $\tan\beta\sim$50,
we see that the enhancement overcomes the loop suppression factor:
\bq\label{aprTS2}
\frac{\delta
  m_{d_i}}{m_{d_i}^0}\simeq \left[10^{-2}\tan\beta\right]
  \left(\frac{\alpha_s}{0.1}\right)\frac{\mu m_{\tilde{g}}}{m_{\tilde{d}_i}^2} .
\eq
If the soft masses are universal, we see that the induced changes are
flavor universal. On the other hand, the needed corrections to the
down and strange quark masses that fix the wrong GUT ratio are far
from universal: If we want to make these corrections in the quark
sector, we must increase the down quark mass while decreasing the
strange mass. Therefore we seem to come to an inevitable situation
that we should depart from the universal soft parameters. Indeed, by
scanning the flavor universal soft SUSY parameter space, several
groups have found no solution to the wrong GUT ratio
\cite{DiazCruz:2000mn, DiazCruz:2005ri, Antusch:2008tf}. In the next subsection, we elaborate on this issue.

Although numerically less
significant due to weaker interactions, the same $\tan\beta$ enhancement occurs to the contributions from
the neutralino-- and  chargino-- squark loops. While they are numerically irrelevant for the lighter two generations, the chargino--stop loop induced correction to the bottom mass could be substantial due to large top Yukawa coupling. In particular, they give
sizeable contributions to the (13) and (23) quark mixings \cite{Blazek:1995nv}. In a certain part of parameter space, such
corrections lead to
$\sim\tan\beta^6$ enhancement for the process
$B^0_s\rightarrow \mu^+\mu^-$ \cite{Babu:1998er}, which could bring it to the
present upper bounds. Furthermore, if a slight discrepancy of $\sim
13$--$24\%$ in the $b$--$\tau$
unification is accounted by these corrections, the branching fraction
of the light Higgs decay to $b\bar{b}$ is altered substantially compared
to the SM or general two--Higgs douplet models.

\subsection{Non--minimal soft A--terms for minimal unification}

Inevitable fact about SUSY extension of the SM is that it inflates the
number of free parameters in the theory from two dozens to over a
hundred upon SUSY breaking in its full generality. They introduce new sources for FCNCs and CP--violations
which have been subject of intensive research for over two
decades. Experimental constraints from $K^0-\bar{K^0}$ mixing, flavor violating $\mu\rightarrow e\gamma$
and many other processes suggest that SUSY breaking parameters are
either flavor blind or aligned with the Standard Model flavor
structure at an extremely high degree. Indeed most of the
phenomenological studies, in particular collider analysis, concentrate on one of the universal
SUSY breaking scenarios. This is certainly understandable considering
the enormous size of the SUSY parameter space, which makes any attempt of generic study
impractical. Secondly there is no compelling theoretical reason
that points to a certain part of parameter space which differs from those
universal ones. On the other hand, as we mentioned earlier, if one wishes to correct the wrong
GUT ratios for the light fermions using the SUSY threshold
corrections, one should most likely depart from a universal choice for soft
terms. Such flavor non--universality in the first two generation could be
phenomenologically quite distinct from the traditional universal scenarios and could lead
to interesting rare decays. If sfermions are discovered at LHC,
their spectrum could confirm/rule out this scenario.

Here first we quantify what amount of threshold corrections are needed for
the minimal $SU(5)$ unification using
approximate expressions. Then discuss the choices of the soft
SUSY breaking parameters  which would induce such corrections.

If one sets the strange quark mass equal to the muon mass at GUT
scale,  without taking into account the threshold effects, its low
energy value would be around $\sim 200$ MeV which is greater by
factor of $\sim$4 than its experimentally measured value. To get the
correct mass value we need corrections of
order $\sim 150$ MeV. The approximate estimate of the gluino--squark
induced term for $s$--quark, for TeV soft masses can be written as follows:
\begin{eqnarray}\label{tsapprox1}
\delta m_s&\simeq&
-\frac{2\alpha_s}{3\pi}v\left( A_s \cos\beta- y_s \mu\sin\beta
\right)\frac{m_{\tilde{g}}}{2m_{\tilde{s}}^2}\nn\\
&\simeq&
\pm25.6\,\mbox{MeV}\left(\frac{\alpha_s}{0.1}\right)\left(\frac{m_{\tilde{g}}}{700\,\mbox{GeV}}\right)\left(\frac{1\,\mbox{TeV}}{m_{\tilde{s}}}\right)^2
\left[5.0\times\left(\frac{A_s}{-500\mbox{GeV}}\right)\left(\frac{10}{\tan\beta}\right)\right.\nn\\
  &&\left. +1.2\times \left(\frac{\mu}{1\,\mbox{TeV}}\right)\left(\frac{y_s}{1.2\times10^{-2}}\right)\right].
\end{eqnarray}
The plus (minus) sign belongs to a positive (negative) value for the
gluino mass parameter. Here we made an approximation
$I(m^2_{\tilde{s}_L},m^2_{\tilde{s}^c},m^2_{\tilde{g}})\simeq
 - 1/(2m^2_{\tilde{s}})$, where $m^2_{\tilde{s}}\sim1$ TeV$^2$. This
crude estimate shows that the $A$-term, if chosen to have a large value, in spite of
$\tan\beta$ suppression, can be quite important and could even become
the dominant source of the threshold
correction. Such large values are subject to the
metastability condition, which will be discuss shortly. To contribute constructively with the $\mu$--term part one
should choose the sign for $A_s$ to be opposite to that of $\mu$.

If $\tan\beta\ga 30$ one approaches the stability with limit
of $|A_s|\la 1.75 \tilde{m}_s$ TeV (See Eq.~(\ref{cond2})). Such a
large value can be easily accommodated by slightly
increasing one of the soft masses in the condition of metastability
given by Eq.~(\ref{cond2}). To reduce the too large value of the $s$
quark mass due to the unification condition, the net effect must give
negative contribution which can be accommodated by the following
choices: (i) positive $m_{\tilde{g}}$ and $A_s$ with preferably
negative $\mu$--term, (ii) negative $m_{\tilde{g}}$ and  $A_s$ with
preferably positive $\mu$--term. This choice lowers the down quark mass which has to be
increased to $m_{d,exp}$. Fortunately one
can see from the discussion of the $s$--quark case, it is much easier
to alter the $d$--quark mass via $A$--term due to its tiny Yukawa
coupling. For example, with similar choice of parameter,
$A_d\sim$ 20--30 GeV is sufficient to induce the needed correction,
which is well within the metastability limit of Eq (\ref{cond2}). If we try to remedy the lighter two
generations only using $\mu$--term, without relying on a
large $A$--term, we must induce $\sim-0.35$ MeV change for the
$d$--quark. Since everything is specified by the unification,
the only freedom left is the choice of the soft masses which translates to
the ratio for their loop functions to be $I_d/I_s\simeq 4.7$ ($I_q\equiv
I(m^2_{{\tilde{q}}_L},m^2_{{\tilde{q}}^c},m^2_{\tilde{g}})$). Such a
large mass splitting in the first two generation is unlikely to survive
severe FCNC constraints for sub TeV scalar masses. In any case, at low and moderate values of $\tan\beta$,
the $\mu$--term part cannot induce enough effects. This leaves us with the
choice of a non--minimal $A$--term for either $d$ or $s$--quark.

Let us consider the case where we correct the $d$--quark mass by a large $A$--term while use
$\mu\tan\beta$ for correcting the $s$--quark mass. To
have  a substantial correction for the $s$--quark
one must choose a larger value for the $\mu\tan\beta$--term. Such a
choice would, at the same time,  reduce
the $b$--quark mass by a potentially large amount. This reduction can not be too large, otherwise it would
jeopardize the already somewhat good $b$--$\tau$
unification. In the MSSM, the RG runnings of the $\tau$ and $b$ Yukawas from
$M_Z$ do not give a unified value at GUT. Instead one obtains
$m_{\tau}(M_{GUT})/m_b(M_{GUT})\simeq 1.13\div1.24$ depending on $\tan\beta$. Since we will
impose the unification condition $m_{\tau}(M_{GUT})/m_b(M_{GUT}) =1$ by
equating the Yukawa coupling of the $b$--quark to that of the $\tau$--lepton, its low
energy value before the threshold correction
will be higher than the experimentally determined value. Thus, to
bring to an agreement one must choose the value of the SUSY breaking
parameters in such a way that the net threshold correction is around $-24\%$ to $-13\%$
depending on the value of $\tan\beta$. Nevertheless this is a much
smaller percentage change compared to the lighter generations: With the same mass as muon at
the GUT scale one gets a factor of $\sim 4$ bigger strange quark mass,
which requires $75\%$ reduction from the radiative
corrections. Therefore, without the effect from large $A$--terms, we have to
choose quite heavier sbottom mass as can be seen from
\bq\label{tsb1}
\frac{\left(\delta m_s/m_s^0\right)}{\left(\delta m_b/m_s^b\right)} \simeq
    \frac{I_s}{I_b}\simeq\frac{m^2_{\tilde{b}}}{m^2_{\tilde{s}}}\simeq
    3\div 5.
\eq
Here we again come to the conclusion that the squark masses have to be
very  different from each other. To summarize, our heuristic arguments
show that when the wrong GUT ratio of the fermion masses are
corrected by the SUSY threshold effects in the down--quark sector, the following options are
available on their soft parameters:

(i) One of the lighter generations has a large $A$--term, while
the sbottom soft masses are heavier than the remaining generations by a factor
of $\sim $ 1.5 to 2.4.

(ii) The $A$--terms are proportional to their Yukawa couplings, while
the soft masses differ from each other by large amount. In this case the
FCNC constraints require them to be very heavy in the range of tens of
TeVs.

(iii) The $s$  and $d$ quarks both have a large $A$--term, where soft
masses can be chosen to be degenerate and not very heavy.

Certainly any of these options are viable. We find the last one
interesting due to its potential implication on LHC
phenomenology, and choose it for our numerical study.

\subsubsection{The metastability condition for $A$--terms}

Beside phenomenological constraints on the soft SUSY breaking parameters there are indirect
constraints coming from the requirement of the stability of our
vacuum \cite{Frere:1983ag}--\cite{Kounnas:1983td}. In Ref.~\cite{Borzumati:1999sp} Borzumati et.~al. have
discussed the implication of this condition  for models of radiative
fermion masses.  To be absolutely stable  against
decaying into color/charge breaking vacua along the $D$--flat
direction where $|f_i|=|f^c_j|=|H|$,
 the following condition must be met the for Yukawa and the trilinear $A$--term,
 $Y_f\tilde{f}^c\tilde{f}H$ and $A_f\tilde{f}^c\tilde{f}H$:
\begin{eqnarray}\label{cond1}
|A_f|\la |y_f|\tilde{m}\nn,\\
\tilde{m}^2\equiv\frac{1}{3}\left(m^2_{\tilde{f}}+m^2_{{\tilde{f}}^c}+m^2_H+\mu^2\right)
\end{eqnarray}
where $m^2_{\tilde{f}}$, $m^2_{\tilde{f}^c}$ and  $m^2_H$ are the soft
masses of the fields connected by the $A$--term. This choice of absolute
stability, which has no physical justification, severely restricts the
size of $A$--term. When applied to the off--diagonal entries it is
more constraining than FCNC processes \cite{Casas:1996de}.

On the other hand if we require only metastability, namely the age of the unstable vacuum is longer than the age of the
Universe, the constraint from numerical analysis gives \cite{Kusenko:1996jn}--\cite{Sarid:1998sn}:
\begin{eqnarray}\label{cond2}
\frac{|A_f|}{\tilde{m}}\la 1.75.
\end{eqnarray}
This metastability
condition allows much more relaxed parameter space compared to the
condition of absolute stability. With a large A--term at one's
disposal it is now much easier to correct light fermion masses by
the SUSY threshold effects. Even for large $\tan\beta$ such a large A--term can easily overcome
$1/\tan\beta$ suppression compared to the $\mu$--term part for the down type fermions and compete
with or even dominate over $\tan\beta$--enhanced contributions. This
is an appealing scenario for any GUT model where one does not need to introduce additional Higgs representations.

\section{Implications of unification through SUSY radiative corrections}

In this section, we present our numerical study of the SUSY radiative
corrections to the fermion masses. In the first subsection we describe
the choice of the soft parameters and the numerical procedures.
In the last part of the section, we discuss the electroweak scale
sfermion spectrum and their experimental implications.

\subsection{The numerical procedures and the results}

Here we describe our numerical calculations and the results. For the
input values of the fermion masses at $M_{SUSY} =$1~TeV we have used
the results of an recent update on the running fermion
masses \cite{Xing:2007fb} (see \cite{Fusaoka:1998vc} for earlier
analysis), where the running masses are calculated in the case of SM
and MSSM  at 1~TeV.
Their results at $1$ TeV for the SM are:
\begin{eqnarray}\label{FM1TeV}
&&m_d=2.50\pm 1.0\,\mbox{MeV},\,\,\,\,m_s=47\pm 14\,\mbox{MeV},\,\,\,\,m_b=2.43\pm0.08\,\mbox{GeV}\nn\\
&&m_u=1.10\pm 0.4\,\mbox{MeV},\,\,\,\,m_c=0.532\pm 0.074\,\mbox{GeV},\,\,\,\,m_t=150.7\pm3.4\mbox{GeV}.
\end{eqnarray}
We take the central values of these results for our analysis. Here we
note that the errors in the estimates given by Xing. et al. in
Ref.~\cite{Xing:2007fb} are from the Particle Data Group
\cite{Amsler:2008zzb} which are notably larger compared to lattice QCD analysis \cite{Blossier:2007vv}. Keeping this in mind we seek
results that are as close to the central values as possible. As for
the  values at $M_{SUSY}=$500~~GeV we have used
\begin{eqnarray}\label{FM500GeV}
&&m_d=2.63\pm 1.05\,\mbox{MeV},\,\,\,\,m_s=51\pm 14\,\mbox{MeV},\,\,\,\,m_b=2.53\pm0.08\,\mbox{GeV}\nn\\
&&m_u=1.17\pm 0.4\,\mbox{MeV},\,\,\,\,m_c=0.553\pm 0.074\,\mbox{GeV},\,\,\,\,m_t=153.6\pm3.4\mbox{GeV}.
\end{eqnarray}

We have used extensively SOFTSUSY
\cite{Allanach:2001kg}, a C++ based publicly available code which
calculates the MSSM spectrum. Initially soft masses are chosen with some
universal values. First we perform MSSM running without the threshold corrections at
SUSY breaking scale $\sim 1$ TeV, to the GUT
scale, $\mu_{GUT}\simeq 2\times10^{16}$ GeV, and impose minimal $SU(5)$ unification:
\begin{eqnarray}\label{unif1}
\label{unifg5}&&\alpha_i=\alpha_U,\,\,\,(i=1,2,3)\\
\label{unifY5}&&Y_5= Y_d=Y_e^T,\\
\label{unifA5}&&A_5= A_e=A_d,\\
\label{unifm5}&&\left(m_{\tilde{L}}^2\right)_{ij}=\left(m_{\tilde{d^c}}^2\right)_{ij}=\left(m_{5}^2\right)\delta_{ij},\\
\label{unifm10}&&\left(m_{\tilde{e^c}}^2\right)_{ij}=\left(m_{\tilde{u^c}}^2\right)_{ij}=\left(m_{\tilde{Q}}^2\right)_{ij}=\left(m_{10}^2\right)\delta_{ij},
\end{eqnarray}
After the GUT conditions in Eq.~(\ref{unif1}) are imposed we run down
back to the SUSY breaking scale. We repeat this enough until we
reach stable low scale values for the Yukawa couplings of the down--type
quarks. These values will tell us then how much
corrections we need from the threshold effects.

We have done the two--loop RGE running to GUT scale and back to
$M_{SUSY}$ to determine the discrepancy between the experimentally
determined values and the values derived from the unification
$y_{d_i}=y_{l_i}|_{GUT}$. Since the effects on the
charged lepton masses are
minor, the initial choice for the leptonic Yukawa couplings before
including the threshold corrections
would be quite close to the full  effective low energy values. Thus,
the numerical entries for the unified Yukawa coupling matrix $Y_5$,
defined in Eq. (\ref{unifY5}), are chosen by the leptonic Yukawa coupling matrix after running them
to GUT scale. Therefore the tree level values of the light
donw--type quark Yukawa couplings will differ from their observed values
significantly leading to the wrong GUT ratio. The objective of our
numerical study is to tackle this problem by identifying the soft
parameters which give the needed corrections. For the $d$ and
$s$--quarks the discrepancies are practically independent of
$\tan\beta$ and we have
\bq\label{delmqlight}
&&\delta m_d \simeq 1.5\,\mbox{MeV and}\,\delta m_s \simeq
-156\,\mbox{MeV
  for}\,\tan\beta=5\div50,\,\,\,M_{SUSY}=1\,\mbox{TeV}.\nn\\
&&\delta m_d \simeq 1.6\,\mbox{MeV and}\,\delta m_s \simeq
-163\,\mbox{MeV for}\,\tan\beta=5\div50,\,\,\,M_{SUSY}=500\,\mbox{GeV}.
\eq
As for the $b$--quark mass, the needed correction mildly depends on
$\tan\beta$. The result
is summarized in Table~\ref{table1}. 
\begin{table}[ht!]
\begin{center}
\begin{tabular}
{|c|c|c|c|c|} \hline
$\tan\beta$                & 5      & 10     & 15     &20  \\ \hline
 $\delta m_b(\mbox{GeV}),\,\,\,M_{SUSY}=1\,\mbox{TeV}$  & -0.690 & -0.699 &  -0.687&-0.666\\
 \hline
$\delta m_b(\mbox{GeV}),\,\,\,M_{SUSY}=500\,\mbox{GeV}$  & -0.710 & -0.720 &  -0.705&-0.693\\ \hline
\end{tabular}
\caption{The needed corrections for the mass of $b$--quark.}
\label{table1}
\end{center}
\end{table}
Since the contribution
from the $A$--term is subleading for $b$--quark, most of
the corrections should come from the $\mu\tan\beta$ part. Then, the
similar change of $\sim$~20$\%$ are automatically induced for the other
generations. For a very large choice of $A_b$ at low $\tan\beta$ one can still get large effect.

\subsubsection{Parameter choices and the induced corrections}

Now we give the details of our choice for the numerical values of the
soft parameters which would yield the needed threshold
effects to give the correct effective masses.

The choices of the initial values of the soft terms, although highly dependent on the SUSY
breaking mechanism and the scale at which it is mediated, we choose
universal scalar masses at GUT scale. We
choose the following parameters as the input of the calculation:

(i) At GUT scale: The trilinear $A$--term for the ${\bf 5}$--plet is
chosen to be  simultaneously diagonalize with its corresponding
Yukawa coupling, $Y_{\bf 5}$, but {\it not} proportional to the
corresponding eigenvalues upon digitalization:
\bq\label{A_5}
\left(A_{\bf 5}\right)_{ij}=a_i\delta_{ij}\neq a_0 y_{\bf 5}.
\eq
The sfermion soft masses are chosen to be flavor--universal
\bq\label{sfmass}
m_{\tilde{\bf 10}}^2=m_{\tilde{Q}}^2\times I_{3\times 3},\nn\\
m_{\tilde{\bf 5}}^2=m_{\tilde{d^c}}^2\times I_{3\times 3},
\eq
For minimal $SU(5)$ unification, we do not have an immediate concern to change the up
sector Yukawa. Therefore, we keep  the trilinear A--term for the ${\bf 10}$  to be minimal at GUT scale $A_{\bf
  10}=a^0_{\bf 10} Y_{\bf 10}$. The gaugino masses are also chosen to be universal.

(ii) The trilinear terms $A_{d}$, $\mu$--term,
the soft highs masses $m^2_{H_u}$ and $m^2_{H_d}$ are chosen at SUSY
breaking scale. We work in the
basis the Yukawa matrices of the charged leptons and down type--quarks
are diagonal. For a chosen set of $ m^2_{\tilde{10}}$ and $ m^2_{\tilde{5}}$,
first we have determined $\mu$ and $m_{1/2}$ that give the correction
given in Table~\ref{table1}.  Samples of these are shown in
Table~\ref{table2} for various choices of $\tan\beta$.
\begin{table}[ht!]
\begin{center}
\begin{tabular}
{|c|c|c|c|c|}\hline
$\tan\beta$   & 5      & 10     & 15     &20
\\ \hline
$m_{1/2}$       &  -210      &  -210     & -230     &-230
 \\ \hline
$m^2_{\tilde{Q}_{i}} \equiv m^2_{\tilde{10}}$      & 0.314      & 0.314
& 0.336     &0.336
\\ \hline
$m^2_{\tilde{d}_{i}} \equiv m^2_{\tilde{5}}$     &0.274    & 0.274     &
0.294     &0.294  \\ \hline
\end{tabular}
\caption{The choices for the soft parameters masses at GUT scale for
  various choices of $\tan\beta$. The units of $m_{1/2}$ is GeV while
  that of the soft masses is TeV$^2$}
\label{table2}
\end{center}
\end{table} 
After this, we have scanned over $A$--terms for the $d$ and
$s$--quarks, $A_d$ and $A_s$, until we obtain the values of the
corresponding mass corrections that are close to the ones given in
Eq.~(\ref{delmqlight}). Such values of $A$--terms are given in
Table~\ref{tabledelm} with the corresponding corrections. As we can
see the desired corrections are obtained. Here we have tried to use
as low values for the $\mu$--term as possible, such that the fine
tuning is minimal. For this reason, $A_b$--term has been chosen to be 
large, except in the case of $\tan\beta=20$ for which the $A$--term
does not contribute significantly without exceeding the stability
condition of Eq.~(\ref{cond2}).
\begin{table}
\begin{center}
\begin{tabular}
{|c|c|c|c|c|}\hline
$\tan\beta$   & 5      & 10     & 15     &20
\\ \hline
$\mu$ (GeV)       & 500  & 550   & 580     &850
 \\ \hline
 $A_d$ (GeV) & 3.5    & 6.4     & 9.2     &16.6 \\ \hline
 $A_s$  (GeV) &  -280    & -460     & -760     &-900   \\ \hline
 $A_b$  (GeV) &  -900    & -950     & -800     &-228     \\ \hline
\hline
$\delta m_{d}$ (MeV)      & 1.50      & 1.43     &
1.55     &1.69  \\ \hline
$\delta m_{s}$ (GeV)      & -0.170      & -0.167     &
-0.158     &-0.156   \\ \hline
$\delta m_{b}$ (GeV)      & -0.730      & -0.732     &
-0.697     &-1.0     \\ \hline
\end{tabular}
\caption{The choices for the $\mu$--term and relevant soft trilinear
  $A$--terms at low energy and the induced change to the down--type quark masses.}
\label{tabledelm}
\end{center}
\end{table}

\subsubsection{The spectrum and its implications}

Due to the large $A$--terms in the second generation, $A_s$ and $A_\mu$, that we found for
the Yukawa unification, the mass degeneracy in the
sfermions of the first two families is lost during the RG running from the GUT scale down to the electroweak scale. If sparticles are eventually discovered at LHC or ILC, this
feature of the mass spectra makes our approach experimentally distinguishable from other
scenarios. At the same time, the induced
splitting cannot be too large, otherwise could exceed the experimental
constraints from the meson oscillations and other rare processes. The
sfermion masses for the above choices of parameters are given in
Table~\ref{table4} and we can see a clear splitting in the masses of
the first two generations. On the other hand, in the present case such a splitting would
be absent in the right up--squark sector, since $A$--terms in the up
sector are chosen to be proportional to the corresponding
Yukawas. This would change if there were large corrections in the up
sector as well. 

\begin{table}
\begin{center}
\begin{tabular}
{|c|c|c|c|c|c|c|c|c|c|}\hline
$\tan\beta$   & 5      & 10     & 15     &20 & $\tan\beta$   & 5      & 10     & 15     &20
\\ \hline
 $m^2_{\tilde{Q}_{1}} $       &   0.505   & 0.507      &
0.569     &0.565  &
$m^2_{\tilde{L}_{1}} $       &   0.280   & 0.375      &
0.295     &0.303  \\ \hline
 $m^2_{\tilde{Q}_{2}} $       &  0.493    & 0.475      &
0.481     &0.440  &
$m^2_{\tilde{L}_{2}} $       &  0.275   & 0.260      &
0.253     &0.243  \\ \hline
 $m^2_{\tilde{Q}_{3}} $       &  0.233    & 0.206      &
0.276     &0.395   &
 $m^2_{\tilde{L}_{3}} $       &  0.202    & 0.175      &
0.198     &0.290 \\ \hline
 $m^2_{{\tilde{d}}^c_{1}} $       &   0.454    & 0.456     &
0.515    &0.508  &
$m^2_{{\tilde{e}}^c_{1}} $    &   0.339   & 0.348      &
0.378     &0.362  \\ \hline
 $m^2_{{\tilde{d}}^c_{2}} $       & 0.431   &0.394   &
0.338     &0.257   &
 $m^2_{{\tilde{e}}^c_{2}} $    &   0.328  & 0.318     &
0.294     &0.241  \\ \hline
 $m^2_{{\tilde{d}}^c_{3}} $       & 0.177      & 0.114     &
0.207     &0.448   &
 $m^2_{{\tilde{e}}^c_{3}} $    & 0.184     & 0.147      &
0.184     &0.337   \\ \hline
\end{tabular}
\caption{The soft sfermion mass parameters at low energy in units of TeV$^2$.}
\label{table4}
\end{center}
\end{table}
The masses in Table~\ref{table4} are calculated in the basis where the down--type
Yukawa coupling matrix is diagonal. Therefore, upon the electroweak
symmetry breaking one must rotate the left--handed up--squark mass matrix by the
CKM matrix. Then, because of the induced splitting $\delta_{12} m^2_{\tilde{Q}}  = m^2_{\tilde{Q}_{1}}-m^2_{\tilde{Q}_{2}}$, the Cabibbo part of the rotation will induce non zero
(12) entry of order $\simeq\lambda\delta_{12} m^2_{\tilde{Q}} $ in the
left up--squark mass matrix. The
immediate consequences are the appearance of non zero mass splittings in neutral $D$--meson and Kaon
systems. The $D_0$--$\bar{D}_0$ oscillation is induced by gluino--up squark
box diagrams while the
$K_0$--$\bar{K}_0$ oscillation is by that of chargino--down
squark box. Observe that there is no gluino box diagram for down sector
at the leading order,
since we have chosen the matrices for the $A$--terms and Yukawa couplings for the down sector to be
simultaneously diagonalize exactly for this reasonto avoid this correction. We have calculated these oscillation
rates and found them at the safe level when checked against the latest
experimental results shown in Table~\ref{table7}. We have included
here also the results for $B$ and $B_s$ meson and are found to be negligible due to the
smallness of the $(13)$, $(23)$ CKM mixings and the chargino--stop
contribution at low and moderate $\tan\beta$'s. As we can see the
effects are large and , in the example of $\tan\beta=20$, there is
already a tension with the experimental result of $K_0$--$\bar{K}_0$. 
\begin{table}
\begin{center}
\begin{tabular}
{|c|c|c|c|c|} \hline
$\tan\beta$              & 5      & 10     & 15     &20  \\ \hline
$\Delta M_{D}\times10^{14} $ GeV$^{-1}$      & 1.44$\times$ 10$^{-2}$      & 0.120     &
0.59     &1.42     \\ \hline
$\Delta M_{K}\times10^{15} $ GeV$^{-1}$      & 1.59$\times$ 10$^{-2}$      & 0.105     &
0.706     &1.55      \\ \hline
$\Delta M_{B}\times10^{15} $ GeV$^{-1}$      & 1.11      & 6.15     &
4.53     &2.93     \\ \hline
$\Delta M_{B_s}\times10^{14} $ GeV$^{-1}$      & 1.82      & 4.56     &
10.1     &45.3     \\ \hline
\end{tabular}
\caption{The mass splittings in $K$ and $B$ meson systems due to the
  SUSY effects.}
\label{table5}
\end{center}
\end{table}

\begin{table}
\begin{center}
\begin{tabular}
{c|c} \hline
$\Delta M_{D}$   & $(1.57\pm^{0.438}_{0.471})\times10^{-14}$ GeV  \cite{Barberio:2008fa}
\\ \hline
$\Delta M_{K}$   & $(3.483\pm0.033)\times10^{-15}$ GeV \cite{Amsler:2008zzb}    \\ \hline
$\Delta M_{B}$   & $(3.337\pm0.006)\times10^{-13}$ GeV \cite{CDF}     \\ \hline
$\Delta M_{B_s}$ & $(1.17\pm0.008)\times10^{-11}$ GeV  \cite{Barberio:2008fa}   \\ \hline
\end{tabular}
\caption{The experimental values of the mass splittings of the neutral
meson.}
\label{table7}
\end{center}
\end{table}

The low energy experimental searches of rare processes in the charm and
strange sectors -- as they are affected most by the non universality
-- could shed light on the spectrum we have obtained. We have examined the rates of $D$ and $K$ meson decays
into pion, neutrino and antineutrino. The process $D^+\rightarrow
\pi^+\bar{\nu}\nu$ has a very tiny branching ratio in the SM for both
short and long distance at the level $\sim 10^{-15}$. Although current upper
bound is still very poor, this will improve soon once BESIII experiment
starts collecting data soon, which would reach to the level of $\sim
10^{-8}$. For our scenario the level has been calculated and found to
be $10^{-11}$ as shown in Table~\ref{table6}. Eventhough four order of
magnitude large than the SM prediction, the rate is beyond the reach
of BESIII. Therefore, further experimental advances are needed regarding this channel.

The processes $K^+\rightarrow \pi^+\bar{\nu}\nu$ and $K^0_L\rightarrow
\pi^0\bar{\nu}\nu$ can be estimated reliably in the SM. The branching
ratio of the process $K^+\rightarrow \pi^+\bar{\nu}\nu$ has been
measured by E787 and E949 collaborations and found to be
$({15.7\pm}^{17.5}_{8.2}) 10^{-11}$ while for the latter there is only
upper bound $6.7\times10^{-8}$. If the data is improved, they provide
a clean probe to possible new physics. Our
results for the $K^+$ decay are shown in Table~\ref{table6}. Indeed in some of our fits we
find quite a large effects from our $A$--term induced splitting. For
example, for $\tan\beta=15$, the result is somewhat larger than the
experimental result. These are
calculated taking into account only the SUSY contributions to see the
effect other than that of the SM. Since we
do not need a complex phase in any of the $A$--terms, the effect cannot be
significant  for the neutral Kaon case. Thus in this case, we expect
the rate to be far from the model independent
Grossman-Nir bound \cite{Grossman:1997sk}.

\begin{table}
\begin{center}
\begin{tabular}
{c|c|c|c|c} \hline
$\tan\beta$              & 5     & 10     & 15     &20  \\ \hline
$Br(D^+\rightarrow \pi^+\bar{\nu}\nu)\times10^{11} $       & 0.0259      & 0.128     &
8.39     & 0.699    \\ \hline
$Br(K^+\rightarrow \pi^+\bar{\nu}\nu)\times10^{11} $      & 0.141      & 3.73     &
37.8     & 19.3    \\ \hline
\end{tabular}
\caption{The branching ratios for processes $D^+\rightarrow
  \pi^+\bar{\nu}\nu$ and $K^+\rightarrow \pi^+\bar{\nu}\nu$.}
\label{table6}
\end{center}
\end{table}

Although we do not have an explicit model which explains the neutrino
oscillation phenomena, most GUT models accommodate the neutrino masses
through the see--saw mechanism, wherein one assumes three standard-model
singlet right--handed neutrinos with masses in the range $\sim
10^{9}\div10^{14}$ GeV. If one of them has a large Yukawa coupling
$y_{\nu_\tau}$ for $y_{\nu_\tau} L_3\nu^cH_u$ type of interaction, it
could alter $\tau$--lepton Yukawa coupling significantly during the
running between its mass and the GUT scale  \cite{Vissani:1994fy,
  Brignole:1994mh}. The $\beta$--function of the $\tau$ Yukawa is
given by:
\bq\label{betaytau}
\mu\frac{dy_\tau}{d\mu}=\beta\left(y_\tau\right)_{MSSM}+\frac{1}{16\pi^2}y_{\tau}
y_{\nu_{\tau}}^2\,\mbox{ for }\,m_{\nu^c_{\tau}}\leq \mu\leq M_{GUT}.
\eq
The additional term increases the value of $y_\tau $ at GUT scale, therefore, increases
$y_b$ compared to $y_\tau $ at low energy scale. Although we have not include this possibility in our analysis,
we would like to point out that if we get a little lower value for
$y_b$ it might be still compatible with the experiment in some specific models with
large $y_{\nu_{\tau}}$. The result for $\tan\beta=20$, shown in Table~\ref{table6}, is one possible
example at hand. In this case if the right--handed neutrino effect
causes $8$--$10\%$ upward deflection on the tau Yukawa coupling
running, the resulting low energy $b$--quark mass would come out
correct.

\section{Conclusions}

In this paper, we have studied the possibilities of Yukawa unification
for all generations in SUSY $SU(5)$ through the finite radiative SUSY
threshold corrections in the presence of flavor non--universal soft parameters.  In
particular, we have concentrated on the flavor non universal $A$--terms and their effect on the Yukawa
couplings of lighter generation fermions. It is well known that these corrections are important and
they can substantially affect the tree level ratios between the third generation
Yukawa couplings. Choosing  the $A$--terms of diagonal form with large values, especially
in the down quark sector, we have shown that the $SU(5)$ unification
with minimal Higgs content for the Yukawa interactions for all
generation is possible for a relatively light sub TeV sfermion
spectrum. This is a welcome scenario at the dawn of the LHC. We have
examined neutral $D$ and $K$ meson mass splittings and rare processes
$D^+\rightarrow \pi^+\nu\bar{\nu}$ and $K^+\rightarrow
\pi^+\nu\bar{\nu}$. While the $D^+$--meson decay has been found to be much more enhanced
compared to the SM, it is still far from the next generation experiments. On
the other hand, we find the latter decay could have a sizable SUSY
contribution and could be probed soon.

In the present scenario, due to the large $A$--term for the second
generation, the down sector squark masses are altered from its
universal value substantially making them distinct from the most
widely studied scenarios. This fact could soon be checked at the
LHC/ILC. When the unification is assumed at the GUT scale for the soft
terms, the slepton spectrum will display a similar distortions as
well. A collider analysis of these type of spectra are needed to be
done thoroughly.

One of the most important questions we have not addressed in the
present work is the problem of the proton decay, which is known to be a
challenge in the minimal $SU(5)$ \cite{Goto:1998qg, Murayama:2001ur}. When the SUSY corrections are used in a
specific GUT models, one will face the question of the proton
decay. One could potentially evade it by the mass splitting between the triplet
and octet which alters the unification scale with a
non--renormalizable operator
\cite{Bajc:2002bv}. Our primary motivation was to avoid exactly these
type of operators. Although this seems to be a setback, at least the
flavor part is not necessarily affected by such operator if used only
for rasing the unification scale. These are currently under investigation.

If experimental breakthroughs happen in both the collider and rare decay experiments, the
sfermion spectrum and rare Kaon decays could point in the direction of
the large $A$--term scenario we have advocated in the present work. We
conclude by noting the fact that in the absence of any theoretical prejudice
for any particular SUSY breaking, all the phenomenologically consistent parameter
space cannot be studied in full generality due to its huge
size. Instead, the ones favored by the unification, such as the minimal
Yukawa unification, could direct us to the parts other than the
universal ones. If SUSY is the solution chosen by the Nature for
stabilizing the gauge hierarchy we may find out soon which one is the
correct one.

\section*{Acknowledgements}

The author would like to thank Goran Senjanovic and Borut Bajc for many
fruitful discussions, Kerim Suruliz for helping with the numerical
calculations, and prof. Kaladi S. Babu at Oklahoma Sate University,
OK, USA and the theoretical physics group of Jozef Stefan Institute, Ljubljana, Slovenia, for giving
the opportunity to present part of the work.

\begin{appendix}
\section{Appendix}
Her we compile the formulae we have used from various sources for our
numerical calculations. These are based on the results of
Refs.~\cite{Rosiek:1989rs, Buras:2002vd}.
The superpotential of the MSSM is given by
\begin{eqnarray}\label{superP1}
W&=&Y^{ij}_u  u^c_iH_uQ_j+Y^{ij}_d  d^c_iH_dQ_j+Y^{ij}_l  l^c_iH_dL_j
+\mu H_u H_d.\nn
\end{eqnarray}
The soft SUSY breaking part of the Lagrangian is given by:
\begin{eqnarray}\label{interaction1}
{\cal -L}&=&
\left(m^2_{L_f}\right)_{ij}\tilde{f}^*_{L_i}\tilde{f}_{L_j}+\left(m^2_{R_f}\right)_{ij}\tilde{f}^{c*}_{i}\tilde{f}^{c}_{j}+\left(m^2_{LR_f}\right)_{ij}\tilde{f}^*_{L_i}\tilde{f}^c_{j}\\
&+&\left(A^{ij}_u  \tilde{u}^c_iH_u\tilde{Q}_j+A^{ij}_d  \tilde{d}^c_iH_d\tilde{Q}_j+A^{ij}_l  {\tilde{l}}^c_iH_d\tilde{L}_j+B\mu H_u H_d\nn\right.\\
&&\left.+\frac{1}{2}m_{\tilde{g}}
\tilde{g}\tilde{g}+\frac{1}{2}m_{\tilde{\chi}^0_a}
\tilde{\chi}^0_a\tilde{\chi}^0_a +\frac{1}{2}m_{\tilde{\chi}^+_a} \tilde{\chi}^+_a\tilde{\chi}^-_a+(\mbox{h.c.})\right).
\end{eqnarray}
With the above conventions, the interactions essential to our analysis are given as:
\begin{eqnarray}\label{interaction2}
{\cal -L}&=&\bar{f}_i\left(N^{L(f)}_{iax}P_L+N^{R(f)}_{iax}P_R\right)\tilde{\chi}^0_a\tilde{f}_x\nn\\
&+&\left(\bar{u}_i\left(C^{L(u)}_{iax}P_L+C^{R(u)}_{iax}P_R\right)\tilde{\chi}^+_a\tilde{d}_x
+\bar{d}_i\left(C^{L(d)}_{iax}P_L+C^{R(d)}_{iax}P_R\right)\tilde{\chi}^-_a\tilde{u}_x\right.\nn\\
&+&\left.\bar{l}_i\left(C^{L(l)}_{iax}P_L+C^{R(l)}_{iax}P_R\right)\tilde{\chi}^-_a\tilde{\nu}_x
+\left(X^{1(f)}_{xy}H^0_d+X^{2(f)}_{xy}H^0_u\right)\tilde{f}_x^*\tilde{f}_y\right.\nn\\
&+&\left.\overline{\tilde{\chi}^0_a}\left(\lambda^{1(N)}_{ab}H^0_dP_L+\lambda^{2(N)}_{ab}H^0_uP_L\right)\tilde{\chi}^0_b
+\overline{\tilde{\chi}^-_a}\left(\lambda^{1(C)}_{ab}H^0_dP_R+\lambda^{2(C)}_{ab}H^0_uP_R\right)\tilde{\chi}^-_b\right.\nn\\
&+&\left.(\mbox{h.c.})\right).
\end{eqnarray}
Here, the neutralino--fermion--sfermion and chargino--fermion--sfermion couplings are given by:
\begin{eqnarray}\label{Couplings1}
N^{L(f)}_{ixa}&=&-\sqrt{2}g_2\tan\theta_WQ_{f}
\left(O^N\right)_{1a}U^{f*}_{x,i+3}
-Y_f^{ij}\left(O^N\right)_{a^\prime a}U^{f*}_{x,j}\\
N^{R(f)}_{ixa}&=&-\sqrt{2}g_2\left\{\tan\theta_W\left(Q_{f}-
T^3_{f}\right)\left(O^N\right)_{a1}+T^3_{f}\left(O^N\right)_{a2}\right\}
U^{f*}_{x,i}\nn \\ &-& Y_f^{ij}\left(O^N\right)_{a^\prime
a}U^{f*}_{x,j+3},\\
C^{L(l)}_{ixb}&=&Y_l^{ij}\left(O^C_R\right)_{b2}U^{\nu*}_{x,j},\\
C^{R(l)}_{ixb}&=&-g_2\left(O^C_L\right)_{b1}U^{\nu*}_{x,i},\\
C^{L(d)}_{ixb}&=&Y_d^{ij}\left(O^C_R\right)_{b2}U^{u*}_{x,j},\\
C^{R(d)}_{ixb}&=&-g_2\left(O^C_L\right)_{b1}U^{u*}_{x,i+3}+Y_u^{ij}\left(O^C_L\right)_{b
2}U^{u*}_{x,j},\\
C^{L(u)}_{ixb}&=&Y_u^{ij}\left(O^C_R\right)_{b2}U^{d*}_{x,j},\\
C^{R(u)}_{ixb}&=&-g_2\left(O^C_L\right)_{b1}U^{u*}_{x,i+3}+Y_d^{ij}\left(O^C_L\right)_{b
2}U^{d*}_{x,j},
\end{eqnarray}
where the sfermion mixing matrices and the mass eigenstates are defined
as follows:
\bq\label{SfermionMixings}
&&\mbox{diag}(m^2_{\tilde{f}})=U^f {\cal M}^2_f
U^{f\dagger},\,\,\,\tilde{f}_x=U^f_{x,i}\tilde{f}_{L_i}+U^f_{x,i+3}\tilde{f}_{R_i},\,\,\,x=(1\div6),\,\,\,f=u,d,l\\
&&diag(m^2_{\tilde{\nu}})=U^\nu {\cal M}^2_\nu U^{\nu\dagger},\,\,\,\tilde{\nu}_x=U^\nu_{x,i}\tilde{\nu}_{i},\,\,\,x=(1\div3).
\eq
Here the charged sfermion $6\times6$ and sneutrino $3\times3$
mass matrices are:
\begin{eqnarray}\label{SoftMasses}
{\cal M }^2_f &=& \left(
\begin{array}{cc}
m_L^2&m^{2\dagger}_{LR}\\
m^{2}_{LR}&m^2_{R} \\
\end{array} \right)\,\,\,\mbox{and}\,\,\,{\cal M}_\nu^2=m_{\tilde{L}}^2+M^2_Z\cos2\beta/2,\\
m_L^2&=&m_{\tilde{f}_L}^2+ Y_f^\dagger Y_fv^2\{\cos^2\beta,
  \sin^2\beta\}/2+M^2_Z\cos2\beta\left(T^3_f-Q_f\sin^2\theta_W\right),\\
m_R^2&=&m_{\tilde{f}_R}^2+ Y_fY_f^\dagger v^2\{\cos^2\beta,
  \sin^2\beta\}/2-M^2_Z\cos2\beta\left(T^3_f-Q_f\sin^2\theta_W\right),\\
m_{LR}^2&=&A_fv\{\cos\beta,-\sin\beta\}/\sqrt{2}-\mu^* Y_f v\{ \sin\beta,\cos\beta
 \}/\sqrt{2}.
\end{eqnarray}
The matrices $O^N$ and ($O^C_L$, $O^C_R$) diagonalize the neutralino and chargino
mass matrices respectively as follows:
\bq\label{ChNmasses}
&&diag(m_{\tilde{\chi}^0_a})=O^N M_N
(O^N)^T,\,\,\,diag(m_{\tilde{\chi}^-_a})=O^C_R M_C (O^C_L)^\dagger,\\
&&(\tilde{B},\tilde{W}_3,H^0_d,H^0_u)=(O^N)^T\tilde{\chi}^0,\,\,({\tilde{W}^-},
{H^-_d})_L=(O^C_L)^\dagger\tilde{\chi}^-_L,\,\,(\tilde{W}^-,H^-_u)=(O^C_R)^\dagger\tilde{\chi}^-_R.
\eq
The index $a^\prime$ of $O^N$ in the
neutralino contribution formula in Eq.~(\ref{Couplings1}) takes value of $3(4)$ for
$T^3_{f}=-\frac{1}{2}(\frac{1}{2})$.
The Higgs--sfermion couplings are given as
\begin{eqnarray}\label{Couplings2}
X^{1(f)}_{xy}&=&D^{f}_{xy}\cos\beta+F^{f}_{xy}\cos\beta+A^{ij}_fU^f_{x,i+3}U^{f*}_{y,j}\,\,\,\mbox{for }  f= d,l,\\
X^{2(f)}_{xy}&=&-D^{f}_{xy}\sin\beta-\mu Y^{ij}_fU^f_{x,i+3}U^{f*}_{y,j}\,\,\,\mbox{for }  f= d,l,\\
X^{1(u)}_{xy}&=&D^{u}_{xy}\cos\beta-\mu Y^{ij}_uU^u_{x,i+3}U^{u*}_{y,j},\\
X^{2(u)}_{xy}&=&-D^{u}_{xy}\sin\beta+F^{f}_{xy}\sin\beta-A^{ij}_fU^f_{x,i+3}U^{f*}_{y,j},\\
X^{1(\nu)}_{xy}&=&D^{\nu}_{xy}\cos\beta+F^{\nu}_{xy}\cos\beta,\\
X^{2(\nu)}_{xy}&=&-D^{\nu}_{xy}\sin\beta,
\end{eqnarray}
where the $D$ and $F$--term induced couplings are given in terms of the following expressions:
\begin{eqnarray}\label{Couplings3}
D^{f}_{xy}&=&\frac{g_2M_Z}{\sqrt{2}\cos\theta_W}\left(\left(T^3_f-Q_f\sin^2\theta_W\right)U^f_{x,i}U^{f*}_{y,i}
+Q_f\sin^2\theta_WU^f_{x,i+3}U^{f*}_{y,i+3}\right),\\
D^{\nu}_{xy}&=&\frac{g_2M_Z}{2\sqrt{2}\cos\theta_W}U^{\nu}_{i,x}U^{\nu*}_{i,y},\\
F^{f}_{xy}&=&\frac{\sqrt{2}M_W}{g_2}\left((Y_fY_f^{\dagger})_{ij}U^f_{i,x}U^{f*}_{j,y}+
(Y_f^{\dagger}Y_f)_{ij}U^f_{x,i+3}U^{f*}_{y,j+3}\right)
,\\
F^{\nu}_{xy}&=&0.
\end{eqnarray}
The neutralino--Higgs and the chargino--Higgs couplings are given by
\begin{eqnarray}\label{Couplings4}
\lambda^{1(N)}_{ab}&=&\frac{g_2}{\sqrt{2}}\left(-\tan\theta_W\left(O^N\right)_{a1}\left(O^N\right)_{b3}
+\left(O^N\right)_{a2}\left(O^N\right)_{b3}\right),\\
\lambda^{2(N)}_{ab}&=&\frac{g_2}{\sqrt{2}}\left(\tan\theta_W\left(O^N\right)_{a1}\left(O^N\right)_{b4}
-\left(O^N\right)_{a2}\left(O^N\right)_{b4}\right),\\
\lambda^{1(C)}_{ab}&=&g_2\left(O^C_R\right)_{a1}\left(O^C_L\right)_{b2},\\
\lambda^{2(C)}_{ab}&=&g_2\left(O^C_R\right)_{a2}\left(O^C_L\right)_{b1},
\end{eqnarray}

Finally, the effective Lagrangian for the Yukawa interactions are given by
\begin{eqnarray}\label{Yukawa1}
-{\cal L}&=& \sum_{f=d,l}\left[(Y_f+\delta Y_f)_{ij}\bar{f}_if_jH^0_d+(\delta Y^{\prime}_f)_{ij}\bar{f}_if_jH^{0}_u\right]\nn\\
&+&(Y_u+\delta Y^\prime_u)_{ij}\bar{u}_iu_jH^0_u+(\delta
Y_u)_{ij}\bar{u}_iu_jH^{0}_d,
\end{eqnarray}
where the corrections, $\delta Y_f$ and $\delta Y_f^\prime$, have
contributions from the gluino--, neutralino-- and chargino--sfermion loops. These are
written as follows:
\begin{eqnarray}\label{Yukawa2}
\left({\delta Y_q^G}\right)_{ij}&=& -\frac{2\alpha_s}{3\pi}X^{1(q)}_{xy}m_{\tilde{g}}U^{q*}_{x,i+3}U^q_{y,j}
I\left( m^2_{\tilde{f}_x},m^2_{\tilde{f}_y},m_{\tilde{g}}^2\right),\\
\left(\delta Y^{N}_f\right)_{ij}&=&-\frac{N^{L(f)}_{iax}N^{R(f)*}_{jay}}{16\pi^2}X^{1(f)}_{xy}m_{\tilde{\chi}^0_a}
I\left(m^2_{\tilde{f}_x},m^2_{\tilde{f}_y},m_{\tilde{\chi}^0_a}^2\right)\nn\\
&-&\frac{N^{L(f)}_{iax}N^{R(f)*}_{jbx}}{16\pi^2}\lambda^{1(N)}_{ab}m_{\tilde{\chi}^0_a}m_{\tilde{\chi}^0_b}
I\left(m^2_{\tilde{f}_x},m_{\tilde{\chi}^0_a}^2,m_{\tilde{\chi}^0_b}^2\right),\\
\left(\delta Y^{C}_u\right)_{ij}&=&-\frac{C^{L(u)}_{iax}C^{R(u)*}_{jay}}{16\pi^2}X^{1(d)}_{xy}m_{\tilde{\chi}^-_a}
I\left(m^2_{\tilde{d}_x},m^2_{\tilde{d}_y},m_{\tilde{\chi}^-_a}^2\right)\nn\\
&-&\frac{C^{L(u)}_{iax}C^{R(u)*}_{jbx}}{16\pi^2}\lambda^{1(C)}_{ab}m_{\tilde{\chi}^-_a}m_{\tilde{\chi}^-_b}
I\left(m^2_{\tilde{d}_x},m_{\tilde{\chi}^-_a}^2,m_{\tilde{\chi}^-_b}^2\right),\\
\left(\delta Y^{C}_d\right)_{ij}&=&-\frac{C^{L(d)}_{iax}C^{R(d)*}_{jay}}{16\pi^2}X^{1(u)}_{xy}m_{\tilde{\chi}^-_a}
I\left(m^2_{\tilde{u}_x},m^2_{\tilde{u}_y},m_{\tilde{\chi}^-_a}^2\right)\nn\\
&-&\frac{C^{L(d)}_{iax}C^{R(d)*}_{jbx}}{16\pi^2}\lambda^{1(C)}_{ab}m_{\tilde{\chi}^-_a}m_{\tilde{\chi}^-_b}
I\left(m^2_{\tilde{d}_x},m_{\tilde{\chi}^-_a}^2,m_{\tilde{\chi}^-_b}^2\right),\\
\left(\delta Y^{C}_l\right)_{ij}&=&-\frac{C^{L(l)}_{iax}C^{R(l)*}_{jay}}{16\pi^2}X^{1(\nu)}_{xy}m_{\tilde{\chi}^-_a}
I\left(m^2_{\tilde{\nu}_x},m^2_{\tilde{\nu}_y},m_{\tilde{\chi}^-_a}^2\right)\nn\\
&-&\frac{C^{L(l)}_{iax}C^{R(l)*}_{jbx}}{16\pi^2}\lambda^{1(C)}_{ab}m_{\tilde{\chi}^-_a}m_{\tilde{\chi}^-_b}
I\left(m^2_{\tilde{\nu}_x},m_{\tilde{\chi}^-_a}^2,m_{\tilde{\chi}^-_b}^2\right),
\end{eqnarray}
To obtain the corrections $\delta Y^\prime_f$ one should replace
$(X^1,\lambda^1)$ by $(X^2,\lambda^2)$ everywhere. In the above formulae, if one
works in the basis where the down (up) type quark Yukawa matrix $Y_d$ ($Y_u$) to be
diagonal then the up (down) type Yukawa matrix must be chosen as
$Y^{diag}_uV_{ckm}$ ($Y^{diag}_dV^\dagger_{ckm}$). The loop function, $I$, is given by
\begin{eqnarray}\label{f1}
&&I\left(x,y,z\right) =
-\frac{xy\ln(x/y)+yz\ln(y/z)+zx\ln(z/x)}{(x-y)(y-z)(z-x)},
\end{eqnarray}
which has the following behaviors in various limits of its arguments as follows
\bq\label{int2}
I(m^2_1,m^2_2,m^2_3) = \left\{
\begin{array}{rl}
1/(2m^2) & \mbox{for }  m_i\rightarrow m,\\
\frac{1}{m^2}\frac{\ln\beta}{\beta-1} & \mbox{for } m_1 =
0,\,\beta\equiv m^2_2/m^2_3,\\
1/m^2\frac{1-\beta+\beta\ln\beta}{(\beta-1)^2} & \mbox{for } m=m_1=m_2.
\end{array} \right.
\eq
\end{appendix}

\end{document}